\documentclass[english,aip,reprint]{revtex4-1}
\usepackage[T1]{fontenc}
\usepackage[latin9]{inputenc}
\usepackage{textcomp}
\usepackage{amstext}
\usepackage{amsmath}
\usepackage{graphicx}



\usepackage{babel}
\begin{document}
\title{3D Strain Field Reconstruction by Inversion of Dynamical Scattering}
\author{Laura Niermann}
\affiliation{Technische Universit\"at Berlin, Institut f\"ur Physik und Astronomie,
Hardenbergstr. 36A, 10623 Berlin, Germany.}
\email{laura.niermann@tu-berlin.de}

\author{Tore Niermann}
\affiliation{Technische Universit\"at Berlin, Institut f\"ur Physik und Astronomie,
Hardenbergstr. 36A, 10623 Berlin, Germany.}
\author{Chengyu Song}
\affiliation{National Center for Electron Microscopy, The Molecular Foundry, Lawrence
Berkeley National Laboratory, Berkeley, CA, 94720, United States.}
\author{Colin Ophus}
\affiliation{Department of Materials Science and Engineering, Stanford University,
Stanford, CA, 94305, United States.}
\begin{abstract}
Strain governs not only the mechanical response of materials but also
their electronic, optical, and catalytic properties. For this reason,
the measurement of the 3D strain field is crucial for a detailed understanding
and for further developments of material properties through strain
engineering. However, measuring strain variations along the electron
beam direction has remained a major challenge for (scanning-) transmission
electron microscopy (S/TEM). In this article, we present a method
for 3D strain field determination using 4D-STEM. The method is based
on the inversion of dynamical diffraction effects, which occur at
strain field variations along the beam direction. We test the method
against simulated data with a known ground truth and demonstrate its
application to an experimental 4D-STEM dataset from an inclined pseudomorphically
grown $\text{Al}_{0.47}\text{Ga}_{0.53}\text{N}$ layer.
\end{abstract}
\maketitle

The internal strain state governs the electronic \citep{Thompson2006},
optical \citep{Jacobsen2006}, and catalytic \citep{Pingel2018} properties
of materials, making its 3D measurement crucial for strain engineering.
Applications range from virtually all modern CMOS-based field effect
transistors \citep{Cao2023}, over strain rich electrocatalysts \citep{Yang2022},
to buried semiconductor quantum dots, where precisely tuned 3D strain
profiles are essential to avoid fine structure splitting in entangled
photon emission \citep{Huber2018}.

However, so far 3D strain fields have remained largely inaccessible
to state-of-the-art strain measurement techniques in (scanning-) transmission
electron microscopy (S/TEM). Several techniques exist like nano-beam
electron diffraction \citep{Usuda2005,Beche2009,Mahr2021},
dark-field electron holography \citep{Hytch2008,Beche2011}, peak
fitting \citep{Bierwolf1993,Rosenauer1996}, geometric phase analysis
\citep{Hytch2003}, or other analyses of atomic resolution images
\citep{Niermann2011}. Critically, all these techniques assume a strain
field that is constant along the electron beam direction. Most of
these techniques can only measure the strain components within the
lateral plane, perpendicular to the electron beam, while the components
in beam direction remain inaccessible for a measurement along a single
direction. Furthermore, strain measurement techniques at atomic resolution
often require thin specimen of a few ten nanometers at maximum, which
alters the strain within the specimen due to surface relaxation effects
\citep{Jacob1998,Beyer2021}. 

For strain measurements the specimen must be oriented into strongly
diffracting orientations, such as systematic row or zone-axis conditions.
Under these orientations dynamical diffraction is dominant in samples
thick enough to be representative of a bulk structure. For strain
fields varying in beam direction dynamical diffraction results in
a highly non-linear dependence of the measured signals on not only
the strain inhomogeneity and its depth, but also on parameters like
specimen thickness and the beam tilt \citep{Lubk2014,Javon2014,Meissner2019},
further complicating strain analysis with most techniques. In contrast
to electrons, X-rays do not exhibit strong dynamical diffraction effects,
which enables advanced computational techniques like Bragg coherent
diffraction imaging to be used for 3D strain measurements \citep{Pfeifer2006, Cherukara2018}. 

In this article, we introduce an inversion method that determines
the 3D strain field from 4D-STEM data. This inversion method measures
the strain not only in lateral directions, but also in the electron
beam direction. The basic idea of this method is to utilize the rich
information encoded in dynamical diffraction patterns to solve the
inverse problem for the 3D strain field. As input we acquire scanning
convergent beam electron diffraction (SCBED) datasets \citep{Niermann2025},
where the electron probe is scanned over the specimen and convergent
beam electron diffraction (CBED) patterns are acquired at each scan position. 
The beam convergence angle is kept smaller than the Bragg angle to 
avoid disk overlap within the CBED patterns. SCBED is a special 4D-STEM
technique that is especially useful for measuring dynamical diffraction
effects from 3D strain fields \citep{Niermann2024,Otto2025}. The
results obtained by the inversion method are spatial mappings of the
strain field not only in the lateral dimension but also along the
electron beam direction. The method itself is widely applicable, since
these SCBED datasets can be recorded with any 4D-STEM capable microscope
and the numerical reconstruction itself can be performed on modern
desktop computers.

Below, we introduce the reconstruction method in detail, test it against
simulated data with known ground truth, and finally demonstrate the
reconstruction of the strain field variations along the beam direction
on an experimental 4D-STEM dataset from an inclined pseudomorphically
grown $\text{Al}_{0.47}\text{Ga}_{0.53}\text{N}$ layer.

To simplify the description, we consider lateral effects only along
the lateral $x$-direction within this article. The $z$-direction
corresponds to the electron beam direction. Furthermore, we only consider
zeroth order Laue Zone scattering within systematic row diffraction
conditions. Under systematic row conditions, all excited crystal reflections
are oriented along a single direction \citep{DeGraef2003}, which
makes the diffraction pattern essentially one-dimensional. Here, the
systematic row is parallel to the $x$-direction. This geometry suffices for the presented
cases. However, this is no general restriction of this method and
its extension to the $y$-direction, different diffraction geometries,
and a more general set of reflections is straightforward.

For the systematic row geometry, all information of the CBED pattern
can be obtained from an one-dimensional profile (with $q$ being the
coordinate along this profile) through the center of the CBED disks
along the systematic row. We can display the intensities of these
diffraction profiles $I(x,q)$ for each spatial $x$-coordinate in
a two dimensional $qx$-plot \citep{Niermann2025}. Each row of this
$qx$-plot corresponds to the diffraction profile obtained at the
corresponding position $x$. Experimental 4D-STEM datasets can be
reduced to such $qx$-plots $I^{(\mathrm{exp})}(x,q)$ by calculating
profiles along the corresponding directions. Examples for such $qx$-plots
can be seen below, for instance in Fig.~\ref{fig:Experimental-layer}(b).
Fig.~\ref{fig:Experimental-layer}(c) shows exemplarily the CBED
pattern, from which the row at $x=9$~nm in the $qx$-plots of Fig.~\ref{fig:Experimental-layer}(b)
is derived.

The deformation of crystals can be described either by its displacement
field $\vec{u}(\vec{r})$ or by the strain tensor. The strain tensor
is defined as the symmetric part of the Jacobian of the displacement
field, while its antisymmetric part describes an infinitesimal rotation
of the crystal. Nevertheless, we will use the term ``strain'' for
the $z$-derivatives of the $x$-component of the displacement field
in the following:
\[
\epsilon(x,z)=\frac{\partial}{\partial z}u_{x}(x,z).
\]

We reconstruct the strain field $\epsilon(x,z)$ numerically as follows:
We choose a regular rectilinear grid with step sizes $\delta x$ and
$\delta z$ for discretization:
\[
\epsilon_{vw}=\epsilon(v\delta x,w\delta z)
\]
with integer indices $v$ and $w$. Evaluations of the strain field
$\epsilon(x,z)$ at positions in between the grid points are performed
by bilinear interpolation of the values of the surrounding grid points.

For a given specimen thickness $t$, and a given strain field $\epsilon(x,z)$
the expected complex-valued beam amplitudes $\phi_{g}$ for the reflections
$g$ can be calculated within the validity of the column approximation
for each $x$-position and each lateral component $k_{\perp}$ of
the incident beam vector by numerical propagation of the Darwin-Howie-Whelan
(DHW) equations \citep{DeGraef2003,Williams2009}:
\begin{multline}
\frac{\partial}{\partial z}\phi_{g}(z;x,k_{\perp})=\mathrm{i}2\pi\left[s_{g}+g\epsilon(x,z)\right]\phi_{g}(z;x,k_{\perp})+ \\
\mathrm{i}\frac{\pi}{k_{0}}\sum_{g'}U_{g-g'}\phi_{g'}(z;x,k_{\perp}),\label{eq:DHW-1}
\end{multline}
with the excitation error
\[
s_{g}=-\frac{g\cdot(g+2k_{\perp})}{2k_{0}},
\]
the structure factors $U_{g}$ of the investigated material, which
we calculate using absorptive atomic form factors\citep{Weickenmeier1991},
and the vacuum wave number $k_{0}.$ In this work, a 4th order Runge-Kutta
scheme with a step size of $0.2$\,nm was used for this propagation
and no material variations were considered within the reconstructions. 

The intensity of a single reflected beam in the exit plane $z=t$
is given by the modulus square of the beam amplitude
\begin{equation}
I_{g}(x,k_{\perp})=\left\Vert \phi_{g}(z\equiv t;x,k_{\perp})\right\Vert ^{2}.\label{eq:intensities}
\end{equation}
and the intensities $I(x,q)$ observed in the corresponding $qx$-plot
are given by (for a residual beam tilt of $\Theta)$:
\begin{equation}
I(x,q)=\sum_{g}I_{g}(x,q-g+k_{0}\Theta)w(q-g).\label{eq:calc-qx}
\end{equation}
The weighting function $w(q)$ is used to mimic the effects of the
condenser aperture with a semiconvergence angle of $\Theta_{\text{max}}$,
but replaces the hard edges of the aperture by the smooth sigmoid
function $\textrm{sig}(q)$ to avoid artifacts:
\begin{multline*}
w(q)=\mathrm{sig}\left(\beta(k_{0}\Theta_{\mathrm{max}}-q)\right)\mathrm{sig}\left(\beta(k_{0}\Theta_{\mathrm{max}}+q)\right) \\
\text{ \,\,with\,\, }\mathrm{sig}(q)=\frac{1}{1+\exp(-q)},
\end{multline*}
where $\beta$ is a softening parameter for the edges.

Having described the physical forward model, we now turn to the inverse
problem. The goal of our inversion method is the reconstruction of
the strain field (i.e. the values for $\epsilon_{vw}$) from $qx$-plots
$I^{\mathrm{(exp)}}(x,q)$ of experimentally measured data. This can
be achieved by numerical minimization of the loss function $L[\epsilon_{vw}]$
with respect to $\epsilon_{vw}$:
\begin{equation}
L[\epsilon_{vw}]=R[\epsilon_{vw}]+\lambda S[\epsilon_{vw}].\label{eq:loss}
\end{equation}
The residual norm 
\begin{equation}
R[\epsilon_{vw}]=\sum_{gxq}\left[w(q-g)\left(NI_{\vec{g}}(x,q-g+k_{0}\Theta)-I^{\text{(exp)}}(x,q)\right)\right]^{2}\label{eq:residual-norm}
\end{equation}
drives the solution towards a strain field that accurately reproduces
the experimental data. The regularization term 

\[
S[\epsilon_{vw}]=\sum_{vw}\epsilon_{vw}^{2},
\]
weighted by $\lambda$, penalizes non-physical, noisy solutions and
promotes a physically plausible, smooth strain field. The parameter
$N$ adjusts the simulated illumination strength to best match the
experimental data by minimizing $R[\epsilon_{vw}]$. The $\mathcal{L}^{2}$-norm
(Tikhonov regularization) is used as regularization term, motivated
by the expectation that the deformation energy will scale with the
square of the strain. The regularization strength $\lambda$ is chosen
using the L-curve criterion \citep{Hansen2001}. In all cases presented
in this article, the residual norm (\ref{eq:residual-norm}) is calculated
only over the data points shown in the $qx$-plots. 

The strain field was determined by numerically minimizing the loss
function (\ref{eq:loss}) using the Broyden--Fletcher--Goldfarb--Shanno
(BFGS) algorithm with automatically calculated gradients. The minimization
was initialized with a zero strain field $\epsilon_{vw}\equiv0$. 

\begin{figure*}
\includegraphics[scale=0.6]{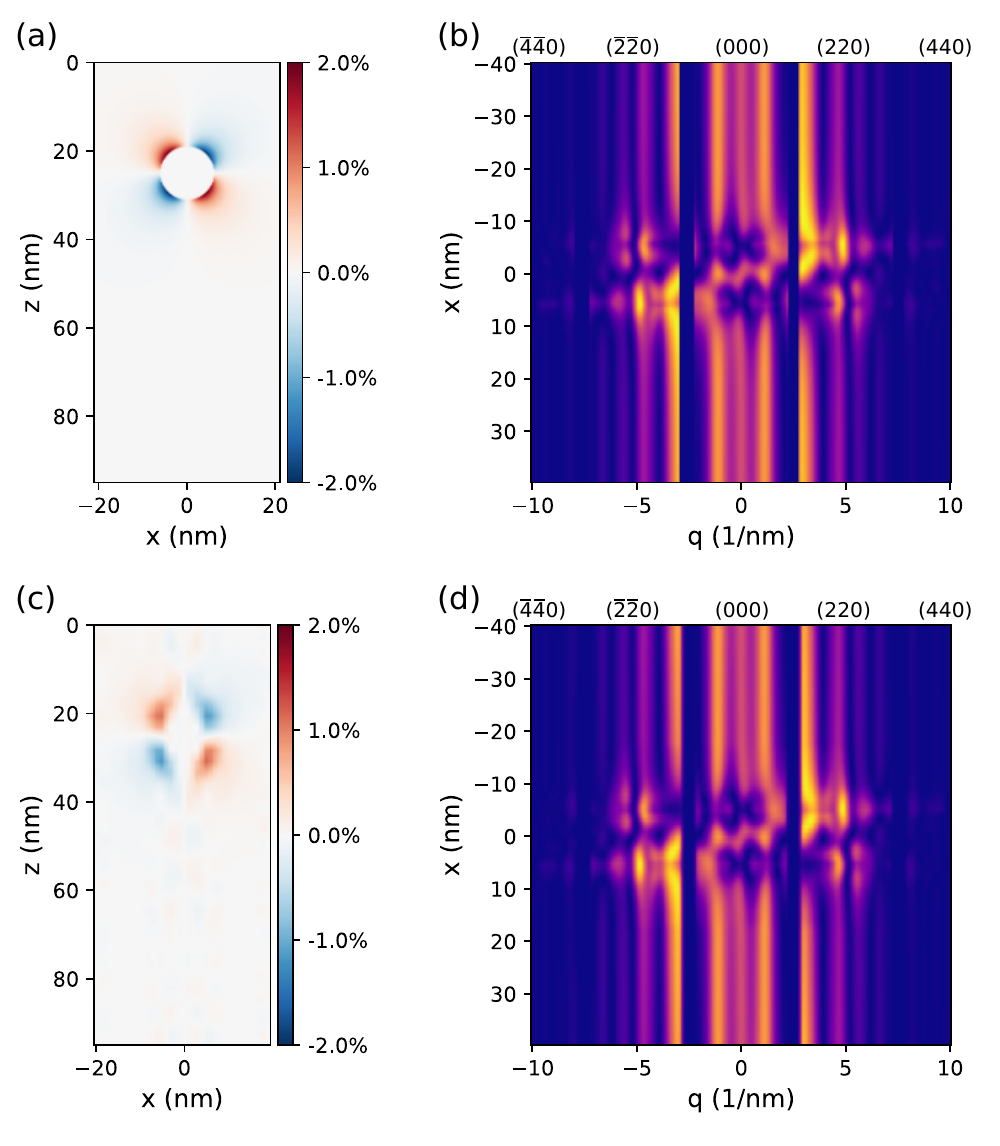}

\caption{(a) Theoretical strain field $\epsilon_{\text{Theo}}(x,z)$ of a spherical
precipitate located at a depth of 25~nm. (b) Simulated $qx$-plot
for the strain field $\epsilon_{\text{Theo}}(x,z)$ in (a). (c) Strain
field $\epsilon(x,z)$ reconstructed by the inversion method from
the simulated $qx$-plot in (b). (d) Simulated $qx$-plot for the
reconstructed strain field $\epsilon(x,z)$ in (c).\label{fig:Spherical-Precipitate}}
\end{figure*}

To first demonstrate the fundamental feasibility of the approach,
we tested it on a simulated dataset with a known ground truth strain
field, namely the 3D strain field of an $\text{In}_{0.4}\textrm{Ga}_{0.6}\textrm{As}$
spherical precipitate, here modeled as the continuum strain model of a hard sphere
\citep{Williams2009}. This test case is an example for specimens
with intrinsic 3D strain fields as not only found for precipitates
but also in buried quantum dots or other inclusions of different materials.
The precipitate has a radius of 6\,nm and is located 25~nm behind
the entrance surface in a 95~nm thick GaAs specimen. The specimen
is oriented in $\{220\}$-systematic row conditions ($g$-direction).
The $x$-direction is parallel to the direction of the systematic
row. The assumed acceleration voltage is 300~kV. Fig.~\ref{fig:Spherical-Precipitate}(a)
shows the theoretical strain field $\epsilon_{\text{Theo}}(x,z)$
(relative to the embedding GaAs). It shows the four characteristic
lobes of the shear strain, which are typical for spherical precipitates
\citep{Williams2009}.

The $qx$-plot shown in Fig.~\ref{fig:Spherical-Precipitate}(b)
was simulated by numerical propagation of the DHW equations for the
theoretical strain field of Fig.~\ref{fig:Spherical-Precipitate}(a)
for a scan across the center of the precipitate. It shows the intensity
variations originating from dynamical diffraction at the strain field
of the precipitate. A semi-convergence angle of 4.3~mrad was used. 

The inversion method is applied to reconstruct the strain field $\epsilon(x,z)$
from this $qx$-plot. The results are shown in
Fig.~\ref{fig:Spherical-Precipitate}(c). The reconstructed strain
shows a lower resolution, especially in the $z$-direction, as can
be seen in the slightly smeared out strain field in beam direction.
Nevertheless, the inversion method has determined the
strain field and the depth of the precipitate solely from the information
encoded within the dynamical diffraction patterns in Fig.~\ref{fig:Spherical-Precipitate}(b).
Fig.~\ref{fig:Spherical-Precipitate}(d) shows the $qx$-plot simulated
for the reconstructed strain, which exhibits a very good match to
the $qx$-plot in Fig.~\ref{fig:Spherical-Precipitate}(b).

The position of the precipitate as well as the four characteristic
lobes are clearly reconstructed. While the position could also be
determined from other methods (e.g. by stereoscopy), the reconstruction
of these four lobes provides insights into the shape of the precipitate,
which were previously not accessible. The actual size of a precipitate
is usually not easily determinable from S/TEM data, as the dominating
contrasts originate from dynamical diffraction of the 3D strain field
and not from the precipitate itself. As strain effects are typically
strongest at the interfaces, the 3D position of the strain features
already give a reliable idea of the precipitate's extents and shape.
Furthermore, stronger localized strain features typically originate
from sharp points in the shape, while extended features like the observed
4 lobes are typical for precipitates with smooth shapes.

\begin{figure*}
\includegraphics[scale=0.6]{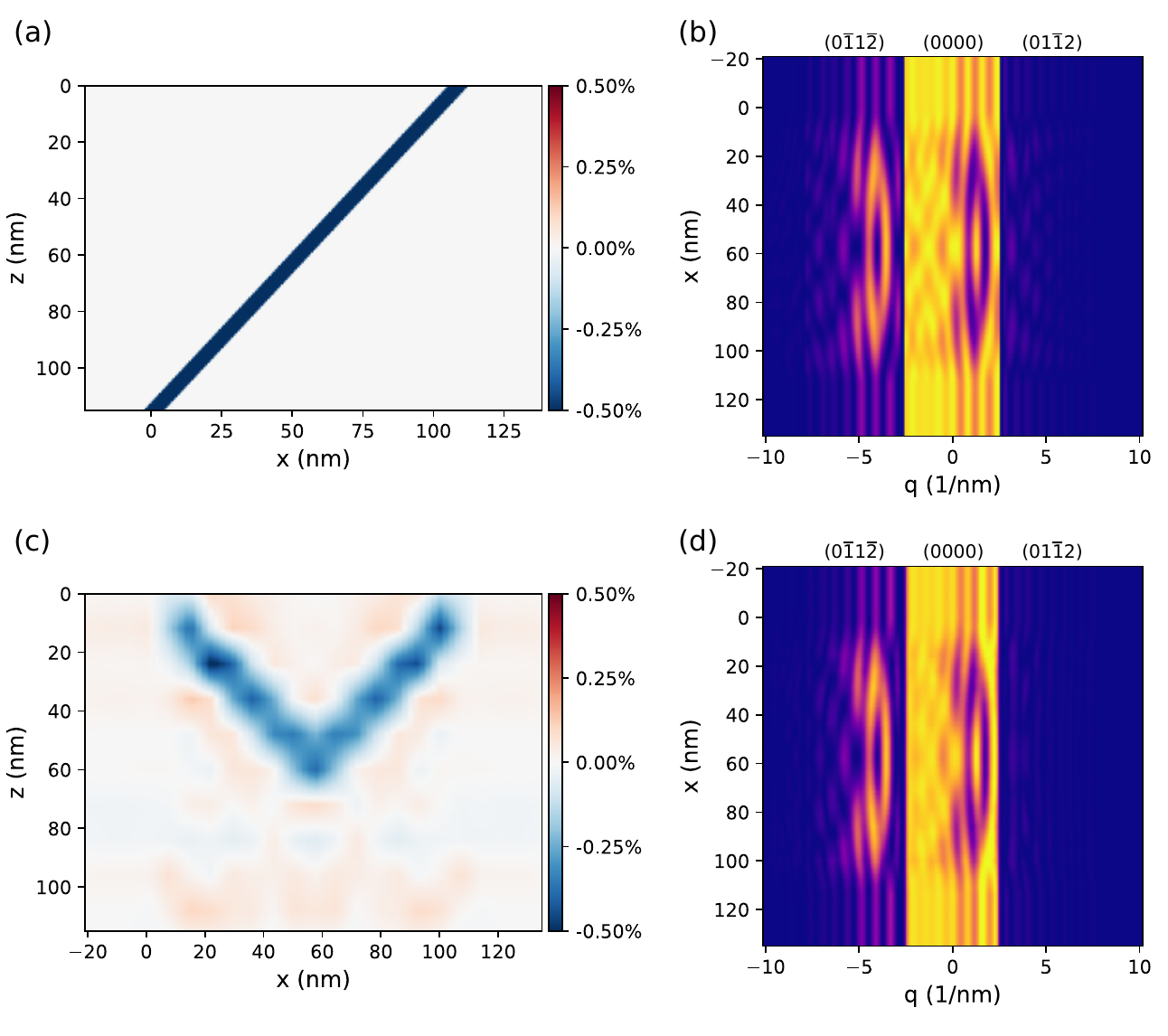}

\caption{(a) Theoretical strain field $\epsilon_{\text{Theo}}(x,z)$ of an
inclined $\mathrm{Al}_{0.47}\mathrm{Ga}_{0.53}\textrm{N}$-layer.
(b) Simulated $qx$-plot for the strain field $\epsilon_{\text{Theo}}(x,z)$
in (a). (c) Strain field $\epsilon(x,z)$ reconstructed by the inversion
method from the simulated $qx$-plot in (b). (d) Simulated $qx$-plot
for the reconstructed strain field $\epsilon(x,z)$ in (c).\label{fig:Simulated-layer}}
\end{figure*}

A key aspect of any dynamical diffraction-based method is understanding
its intrinsic symmetries. To investigate this, we applied the method
to a simulated inclined layer, a geometry known to produce symmetric
diffraction effects. The results are shown in Fig.~\ref{fig:Simulated-layer}.
The assumed specimen contains an inclined $\mathrm{Al}_{0.47}\mathrm{Ga}_{0.53}\textrm{N}$-layer
of 5~nm thickness embedded within a 115~nm thick GaN matrix. This
application is an example for strain fields in specimen with extended
structures as found for instance in semiconductor heterostructures,
but also is prototypical for extended defects like stacking faults.
The $\mathrm{Al}_{0.47}\mathrm{Ga}_{0.53}\textrm{N}$-layer is located
within the basal plane of the Wurtzite structure. The specimen is
oriented in $\{01\overline{1}2\}$-systematic row conditions. The
$x$-direction is parallel to the systematic row direction. The crystal's
$c$-axis lays within the $xz$-plane with an inclination of 43\textdegree\ with
respect to the $x$-direction. The assumed acceleration voltage is
200~kV. Fig.~\ref{fig:Simulated-layer}(a) shows the theoretical
strain field $\epsilon_{\text{Theo}}(x,z)$ of the inclined layer
(relative to the GaN host material). 

The $qx$-plot shown in Fig.~\ref{fig:Simulated-layer}(b) was simulated
by numerical propagation of the DHW equations for the theoretical
strain field of Fig.~\ref{fig:Simulated-layer}(a). A semiconvergence
angle of 6.3~mrad was used. The $qx$-plot in Fig.~\ref{fig:Simulated-layer}(b)
shows a mirror symmetry in $x$-direction with respect to $x\approx67$~nm,
which is the position where the inclined layer crosses the specimen's
midplane. 

The inversion method is applied to reconstruct the strain field $\epsilon(x,z)$
from this $qx$-plot. The resulting reconstructed strain field $\epsilon(x,z)$
is shown in Fig.~\ref{fig:Simulated-layer}(c). The reconstructed
strain field $\epsilon(x,z)$ in Fig.~\ref{fig:Simulated-layer}(c)
and the theoretically strain field $\epsilon_{\text{Theo}}(x,z)$
in Fig.~\ref{fig:Simulated-layer}(a) differ significantly, even
though the corresponding simulated $qx$-plots Fig.~\ref{fig:Simulated-layer}(b)
and Fig.~\ref{fig:Simulated-layer}(d) are nearly identical.

This is a direct consequence of a well-known symmetry of dynamical
diffraction \citep{Koprucki2022,Pogany1968,Howie1961}: a mirroring
of the strain field $\epsilon(x,z)$ with respect to the specimens
midplane $t/2$ results in the same $qx$-plots
\[
I[\epsilon(x,z)]=I[\epsilon(x,t-z)].
\]
This symmetry was also already observed in experimental $qx$-plots
of inclined layers \citep{Niermann2025} and dislocations \citep{Niermann2024}. 

\begin{figure*}
\includegraphics[scale=0.4]{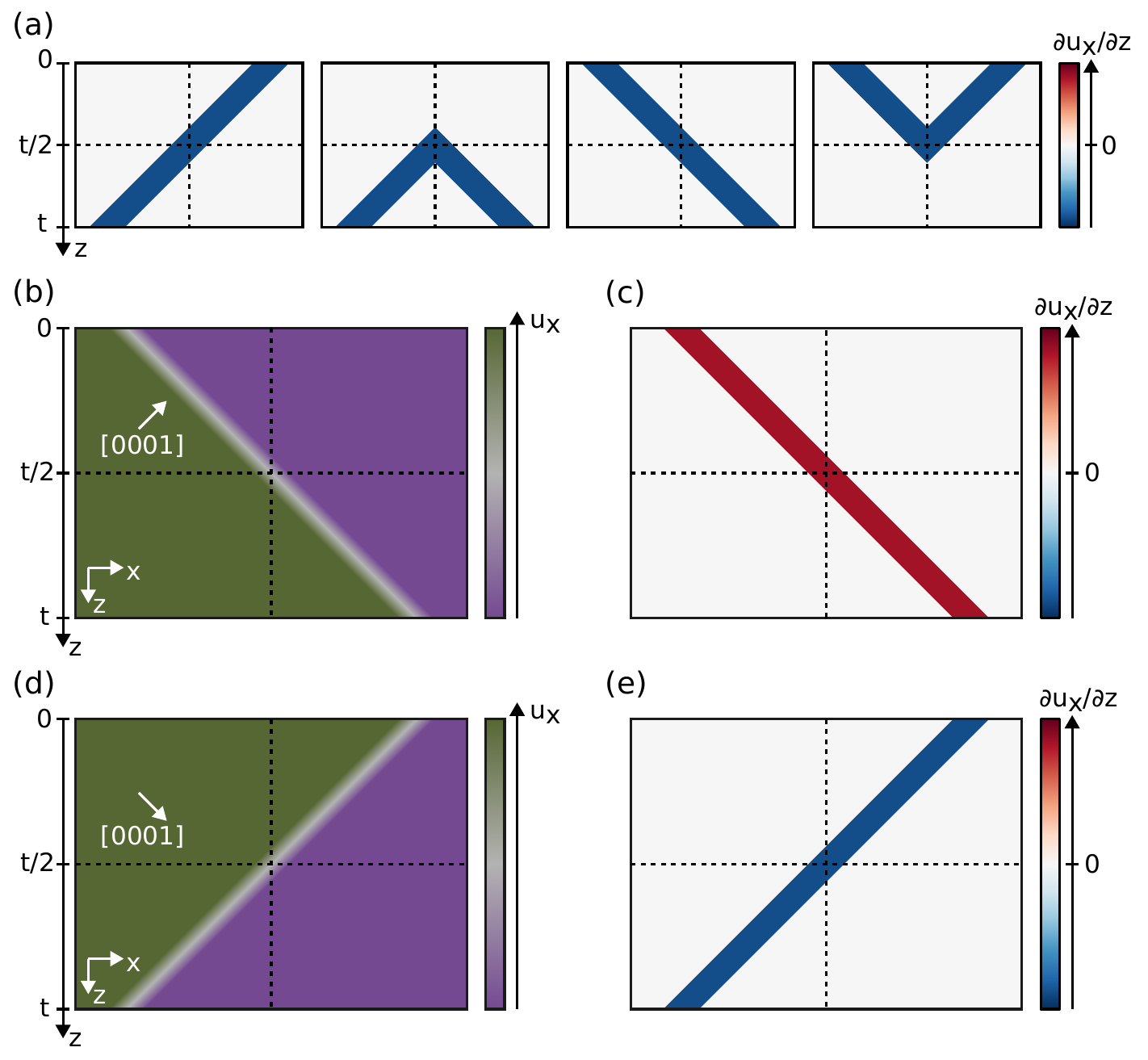}

\caption{(a) Sketches of strain fields $\epsilon(x,z)$, which will result
in the same $qx$-plots. These four strain fields are obtained by
flipping the left half, the right half, or both of them at the specimen
midplane at $z=t/2$. (b) Displacement field $u_{x}$ with $[0001]$-direction
inclined towards the top surface. (c) Derivative $\partial u_{x}/\partial z$
of the displacement field in (b). (d) Displacement field $u_{x}$
with $[0001]$-direction inclined towards the bottom surface. (e)
Derivative $\partial u_{x}/\partial z$ of the displacement field
in (d). \label{fig:Symmetry}}
\end{figure*}

This symmetry leads to an ambiguity in the reconstructed strain profiles.
In principle, this ambiguity might lead to an arbitrary flipping of
the reconstructed $\epsilon(x,z)$ with respect to $t/2$ between
neighboring $x$ points. However, the used strain discretization with
a coarse $x$-spacing stabilizes the reconstructions with a continuous
$\epsilon(x,z)$ in $x$-direction. Both effects combined lead to
the four compatible continuous reconstructions of the strain fields
$\epsilon(x,z)$ that are shown in Fig.~\ref{fig:Symmetry}(a). All
four strain fields will result in nearly the same intensities within
the $qx$-plots, thus are mathematically correct solutions to the
inversion problem. The reconstruction method converges
on one of the possible solutions in Fig.~\ref{fig:Symmetry}(a).
Which of the four cases eventually is reconstructed will depend on
the starting conditions (and in experimental data also on noise and
other experimental influences). 

However, this mathematical ambiguity is not a practical impasse. Only
one of these solutions is typically physically plausible, therefore
a unique solution can almost always be identified through minimal
physical or crystallographic considerations: Since the $\mathrm{Al}_{0.47}\mathrm{Ga}_{0.53}\textrm{N}$-layer
has smaller lattice constants than the embedding GaN, the displacement
field component in the crystalline $[0001]$-direction (the Wurtzite's
$c$-axis) must decrease in this direction as the elongation along
the $c$-axis is negative. However, two possibilities for the orientation
of the basal plane exist, which are compatible with a $\{01\overline{1}2\}$-normal
in $x$-direction. These are shown in Fig.~\ref{fig:Symmetry}(b)
and \ref{fig:Symmetry}(d). Their resulting derivatives of the displacement
field along the $z$-direction are shown in Fig.~\ref{fig:Symmetry}(c)
and \ref{fig:Symmetry}(e). However, only the one in Fig.~\ref{fig:Symmetry}(e)
will exhibit a negative derivative within the layer, making this the
only physically possible solution.

Having tested the method using simulations, we now demonstrate its
application to a real-world experimental dataset as shown in Fig.~\ref{fig:Experimental-layer}.
The experimental data was obtained from a zero-loss filtered 4D-STEM
measurement of an inclined $\mathrm{Al}_{0.47}\mathrm{Ga}_{0.53}\textrm{N}$-layer
of $5.0\ensuremath{\pm}0.5$~nm thickness embedded within a $52.5\ensuremath{\pm}2.5$~nm
thick GaN lamella. This specimen was chosen, as it allows for a controlled
strain field varying in the depth of the specimen. The layer was grown
on the basal plane of the Wurtzite structure and was previously characterized
in on-edge orientation. The inclination was realized by preparation
of a lamella under an angle of 43\textdegree\ with respect to the growth direction
by means of focused ion beam milling \citep{Meissner2019} (see supplementary
material for details). 

The specimen is oriented in $\{01\overline{1}2\}$-systematic row
conditions and the $x$-direction is along the $\{01\overline{1}2\}$-normal.
The crystal's $c$-axis lays within the $xz$-plane with an inclination
of 43\textdegree\ with respect to the $x$-direction. The expected strain field
for this specimen under the assumption of a pseudomorphic growth\citep{Pohl2013}
and without the consideration of surface relaxations is shown in Fig.~\ref{fig:Experimental-layer}(a).
The measurement was performed on TEAM I, a double-aberration-corrected
FEI Titan 80-300 microscope, operating at 300~keV. The 4D-STEM data
was acquired using the GIF camera with a binning of 4 (for further
acquisition details see the supplementary material). 

Fig.~\ref{fig:Experimental-layer}(b) shows the $qx$-plot obtained
from this measurement. The experimental semiconvergence angle was
5.0~mrad, which corresponds to a probe size of roughly 0.4~nm. Clearly
visible is the change within the $qx$-plots for $x$-coordinates
between -25~nm and 25~nm, which is caused from the scattering at
the inclined layer. Furthermore, a slight shifting of the reflections
itself towards negative $q$-coordinates with increasing $x$-coordinates
can be seen within the measurement. This is a measurement artifact
caused by a misalignment of the beam shift-tilt decoupling within
the microscope. Furthermore, a slight shift of the intensity patterns
within the reflections to negative $q$ coordinates with increasing
$x$-coordinates can be noticed. This is caused by a slight bending
of the specimen \citep{Niermann2025}. Both artifacts have been corrected
by incorporating an additional shear transformation within the $qx$-plane
into the reconstruction. The lamella thickness $t=52.5\text{~nm}$
and the residual beam tilt were evaluated in advance by minimizing
the loss function with respect to these parameters in the unstrained
areas. 
\begin{figure*}
\includegraphics[width=17cm]{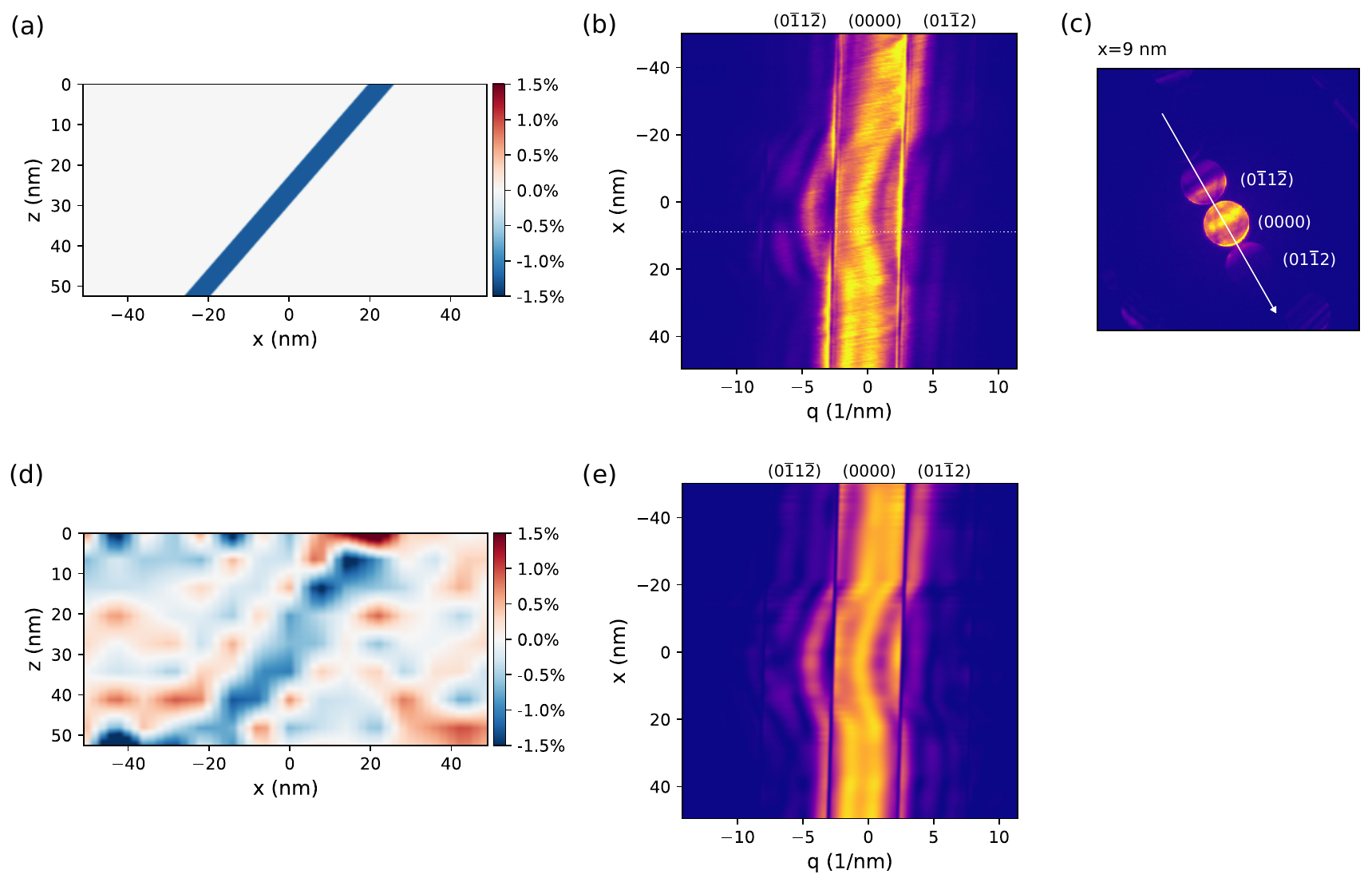}

\caption{(a) Expected strain field for a pseudomorphically grown 5.0~nm thick
$\mathrm{Al}_{0.47}\mathrm{Ga}_{0.53}\textrm{N}$-layer. (b) Experimentally
obtained $qx$-plot from the inclined layer. (c) Exemplary CBED pattern,
with indicated profile direction, which corresponds to the row at
$x=9$~nm indicated in (b). (d) Strain field $\epsilon(x,z)$ reconstructed
from data in (b). (e) Calculated $qx$-plot for reconstructed strain
field in (d).\label{fig:Experimental-layer}}
\end{figure*}

The inversion method is applied to reconstruct the strain field $\epsilon(x,z)$
from this $qx$-plot. The result is shown in Fig.~\ref{fig:Experimental-layer}(d).
Within Fig.~\ref{fig:Experimental-layer}(d) the inclined layer is
the prevailing feature and it well reproduces to the strain field
expected for the layer, which is shown in Fig.~\ref{fig:Experimental-layer}(a).
The reconstructed strain field in Fig.~\ref{fig:Experimental-layer}(d)
represents one of the mathematically valid solutions consistent with
the symmetries discussed above. We identify this as the physically
correct solution based on the known expected compressive strain in
the AlGaN layer. The excellent agreement between the $qx$-plot Fig.~\ref{fig:Experimental-layer}(e)
simulated for the reconstructed strain field and the experimental
data in Fig.~\ref{fig:Experimental-layer}(b) -- despite the presence
of shot noise and minor experimental artifacts -- demonstrates the
practical robustness of the method. 

We demonstrated the reconstruction of 3D strain fields by the presented
reconstruction method. However, it is important to understand its
current limitations and potential sources of error. We identify three
main categories of uncertainty: (i) fundamental ambiguities arising
from physical symmetries, (ii) systematic errors from inaccuracies
in experimental parameters such as specimen thickness and residual
tilt and (iii) statistical uncertainty from experimental noise.

The primary limitation is the inherent ambiguity due to the intrinsic
symmetry of dynamical diffraction. A key direction for future work
will be the implementation of strategies for breaking this symmetry,
removing the need for post-reconstruction analysis and paving the
way for a fully automated 3D strain mapping workflow. One option to
overcome this ambiguity is the use of additional regularizations,
which favors the physically correct solution. Another option is to
evaluate the beam phases, which, in contrast to the intensities, are
not susceptible to the symmetry. This can be achieved either by utilizing
an overlap between neighboring CBED disks like also used in ptychography
or by the acquisition of holographic tilt-series instead of STEM-data.
A third option is the incorporation of additional measurements taken
with rotated specimen into the inversion method. Incorporating multiple
projection directions also will enable the 3D mapping of all components
of the strain tensor. 

We mitigated the influence of systematic errors, by determining these
parameters from strain-free reference regions of the sample prior
to the main reconstruction. The camera length (or more specifically
the reciprocal space sampling of the CBED patterns) as well as the
effects of misaligned decoupling of beam tilt and shift, can be calibrated
from the positions of the reflections in these reference regions, as it is also
standard in state-of-the-art STEM based measurement methods. Residual
tilt and thickness can be estimated by comparing the CBED intensity
patterns in these areas with simulations. The influences of these
systematic effects can also be accounted for within the reconstruction:
we also added the possibility to our method to optimize for the specimen
thickness and reciprocal space sampling in addition to the strain
field. However, we noticed no significant differences in the presented
reconstructions. Residual tilt of the specimen will occur as a constant
background in the reconstructed strain, as infinitesimal rotations
of the specimen contribute to the derivative of the displacement field.
Eventually this is more a question of relative to which reference
the strain is measured. In principle defocus can be accounted for
as an additional shear of the dataset within the $qx$-plane \citep{Niermann2025}. 

As demonstrated in the reconstruction from experimental
data, shot noise is no limitation in the present example, but obviously
will influence the quality of the reconstructions under lower dose
conditions.

Future work will focus on quantifying the method's precision by systematically
investigating the impact of noise and other experimental influences
on the reconstructed strain. The demonstrated performance of the method
under standard acquisition conditions suggest that high-fidelity reconstructions
are achievable.

In conclusion, we have developed and demonstrated an inversion method
that reconstructs 3D strain fields from 4D-STEM data, providing access
to strain variations along the electron beam direction - a capability
unavailable to existing S/TEM techniques. Our technique is not confined
to thin specimens; it is specifically designed for thicker samples
(>50 nm), where it can measure strain variations deep within the material,
free from the surface relaxation effects that challenge conventional
strain measurement techniques. Furthermore, the method does not require
a-priori models of the strain distribution. 

Our method provides a direct solution to the long-standing challenge
of quantifying 3D strain fields in technologically relevant systems,
such as catalysts and (opto-) electronic semiconductor devices. By
inverting dynamical diffraction effects, we can now, for instance,
resolve the size, shape, and strain state of buried features like
quantum dots and precipitates in 3D -- information previously obscured
by the effects of dynamical diffraction. The determination of 3D strain
fields is immediately applicable to optimizing the optoelectronic
properties of semiconductor heterostructures, understanding failure
mechanisms in structural alloys, and engineering the catalytic activity
of core/shell nanoparticles. By providing a quantitative, depth-resolved
picture of strain, this technique enables a more precise correlation
between nanoscale structure and device performance, offering a powerful
tool for materials design and process control.

\section*{Supplementary Material}

See supplementary material for a more detailed description of the
investigated specimen and its preparation, and for further parameters of the experiment
and the reconstruction.

\begin{acknowledgments}
L. N. acknowledges support by the Deutsche Forschungsgemeinschaft
(DFG, German Research Foundation) under Project 492463633. Work at
the Molecular Foundry was supported by the Office of Science, Office
of Basic Energy Sciences, of the U.S. Department of Energy under Contract
No. DE-AC02-05CH11231. We thank the Center for Electron Microscopy
(ZELMI), Technische Universit\"at Berlin, and the Structure Research
and Electron Microscopy group, Humboldt-Universit"at zu Berlin, in
the context of the Alliance Center Electron Microscopy (ACEM) for
support. The ACEM is funded under the Excellence Strategy of the Federal
Government and the L\"ander by the Berlin University Alliance (BUA).
\end{acknowledgments}

\section*{Data Availibility Statement}

The data that support the findings of this study are openly available
in the Zenodo repository at {\url
  {https://doi.org/10.5281/zenodo.15309273}}\citep{InverseQX_Data}

\section*{Author Declarations}

The authors have no conflicts to disclose.

\section*{Author Contributions}

\textbf{Laura Niermann:} Conceptualization, Funding Acquisition,
Investigation, Methodology, Project Administration, Resources, Software,
Supervision, Writing -- original draft, Writing -- review \& editing.
\textbf{Tore Niermann:} Investigation, Visualization, Data Curation,
Software, Validation, Writing -- review \& editing. \textbf{Cheyun Song:}
Investigation, Writing -- review \& editing. \textbf{Colin Ophus:}
Resources, Validation, Writing -- review \& editing.

\bibliographystyle{apsrev4-2}

\pagebreak
\widetext
\begin{center}
\textbf{\large Supplemental Materials: 3D Strain Field Reconstruction by Inversion of Dynamical Scattering}
\end{center}
\setcounter{equation}{0}
\setcounter{figure}{0}
\setcounter{table}{0}
\setcounter{page}{1}
\makeatletter
\renewcommand{\thesection}{S\arabic{section}}
\renewcommand{\theequation}{S\arabic{equation}}
\renewcommand{\thefigure}{S\arabic{figure}}
\renewcommand{\thetable}{S\arabic{table}}
\renewcommand{\bibnumfmt}[1]{[S#1]}
\renewcommand{\citenumfont}[1]{S#1}

\section{Details on experiment}

\begin{figure}[b]
\includegraphics[width=1\textwidth]{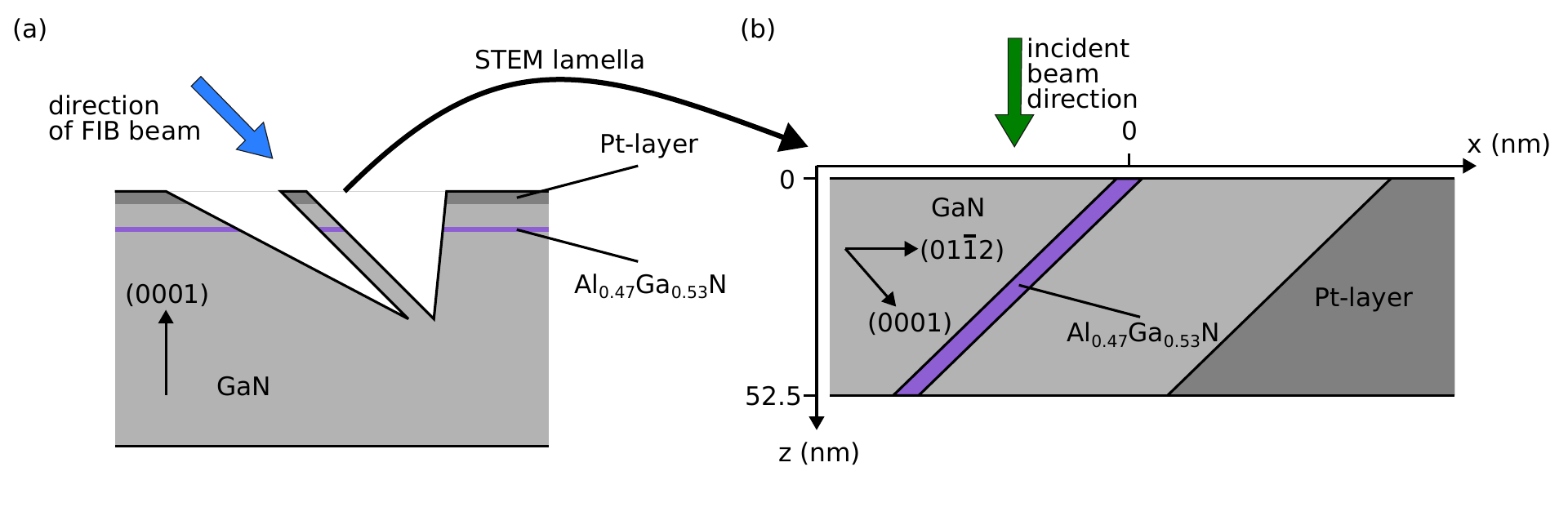}

\caption{FIB preparation of the inclined layer structure: (a) The FIB lamella
preparation was performed under an angle of 43\textdegree\ with respect to the
wafer surface and the $\mathrm{Al}_{0.47}\mathrm{Ga}_{0.53}\textrm{N}$-layer.
(b) The resulting STEM lamella contains the inclined layer. \label{fig:FIB-preparation}}

\end{figure}

The STEM lamella, from which the experimental data was obtained, was
prepared by focused ion beam (FIB) milling from a semiconductor heterostructure.
The heterostructure consists out of a 5~nm thick $\mathrm{Al}_{0.47}\mathrm{Ga}_{0.53}\textrm{N}$-layer
embedded within GaN. The layer structure was grown on (0001) planes
using metal-organic chemical vapor deposition. Thickness and concentration
of the layer were verified by means of EDX-STEM on a separate cross-section
STEM-lamella. For the inclined layer structure investigated in this
work the lamella was prepared under an angle 43\textdegree\ with respect to the
wafer's (0001) surface as sketched in Fig.~\ref{fig:FIB-preparation}(a).

After FIB preparation the lamella was transferred into the STEM, where
the lamella is observed with the lamella surfaces nearly perpendicular
to the beam. The specimen is tilted within the STEM a few degrees
around the $(01\overline{1}2)$-normal to achieve $(01\overline{1}2)$-systematic
row conditions. In this orientation the specimen eventually contains
an inclined $\mathrm{Al}_{0.47}\mathrm{Ga}_{0.53}\textrm{N}$-layer
with a well known geometry as shown in Fig.~\ref{fig:FIB-preparation}(b). 

\begin{figure}
\includegraphics[width=0.7\textwidth]{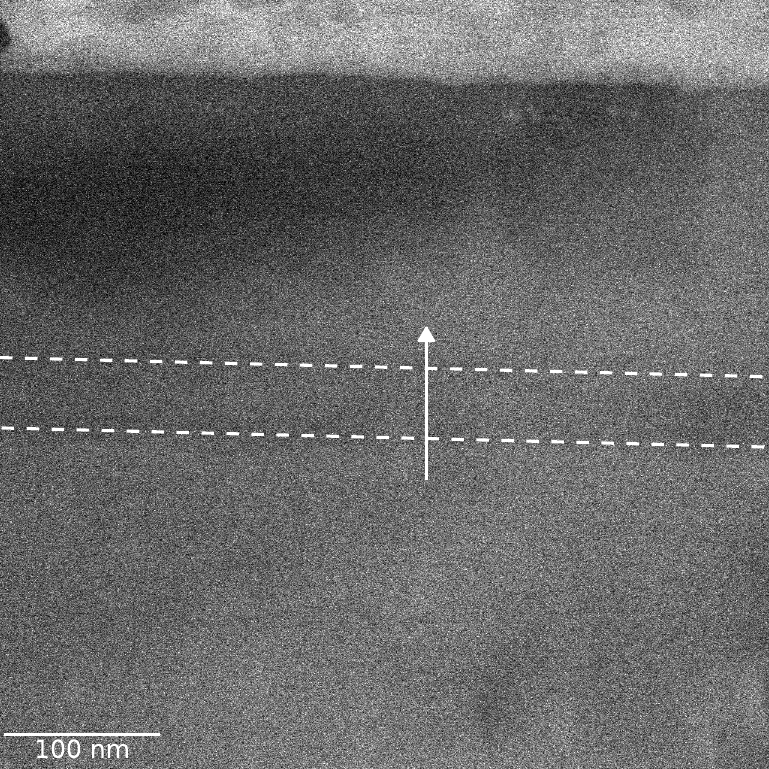}

\caption{Annular dark field overview image over the inclined $\mathrm{Al}_{0.47}\mathrm{Ga}_{0.53}\textrm{N}$-layer.
The dashed lines mark the projected edges of the inclined layer. The
arrow indicates the region over which the profile within the spatial
dimensions was obtained. The position $x$ within the $qx$-plots
corresponds to a point on this profile. In the top part of the image
the protective Pt layer can be seen.\label{fig:ADF overview}}

\end{figure}

The 4D-STEM data was obtained with a scan-stepping of 0.31~nm. The
sampling of the diffraction pattern was 0.11~mrad/px. For the $qx$-plots,
at first profiles in spatial dimensions were obtained across the inclined
layer. Fig.~\ref{fig:ADF overview} shows an annular dark field overview
over the inclined layer. At the top of the image the protective Pt-layer
from FIB preparation is visible. Within the region of the projected
inclined layer the image contrast is slightly decreased. For better
visibility the upper and lower projected edge are indicated by a dashed
line in Fig.~\ref{fig:ADF overview}. The position of the aforementioned
profile is indicated by the arrow. The individual diffraction patterns
of the 4D-STEM dataset were averaged over a window of 16~px width
in the perpendicular direction (parallel to the inclined layer) to
increase the signal-to-noise ratio. This results in a 3D datasets
(an averaged 2D diffraction pattern for each spatial position $x$).

\begin{figure}
\includegraphics[width=1\textwidth]{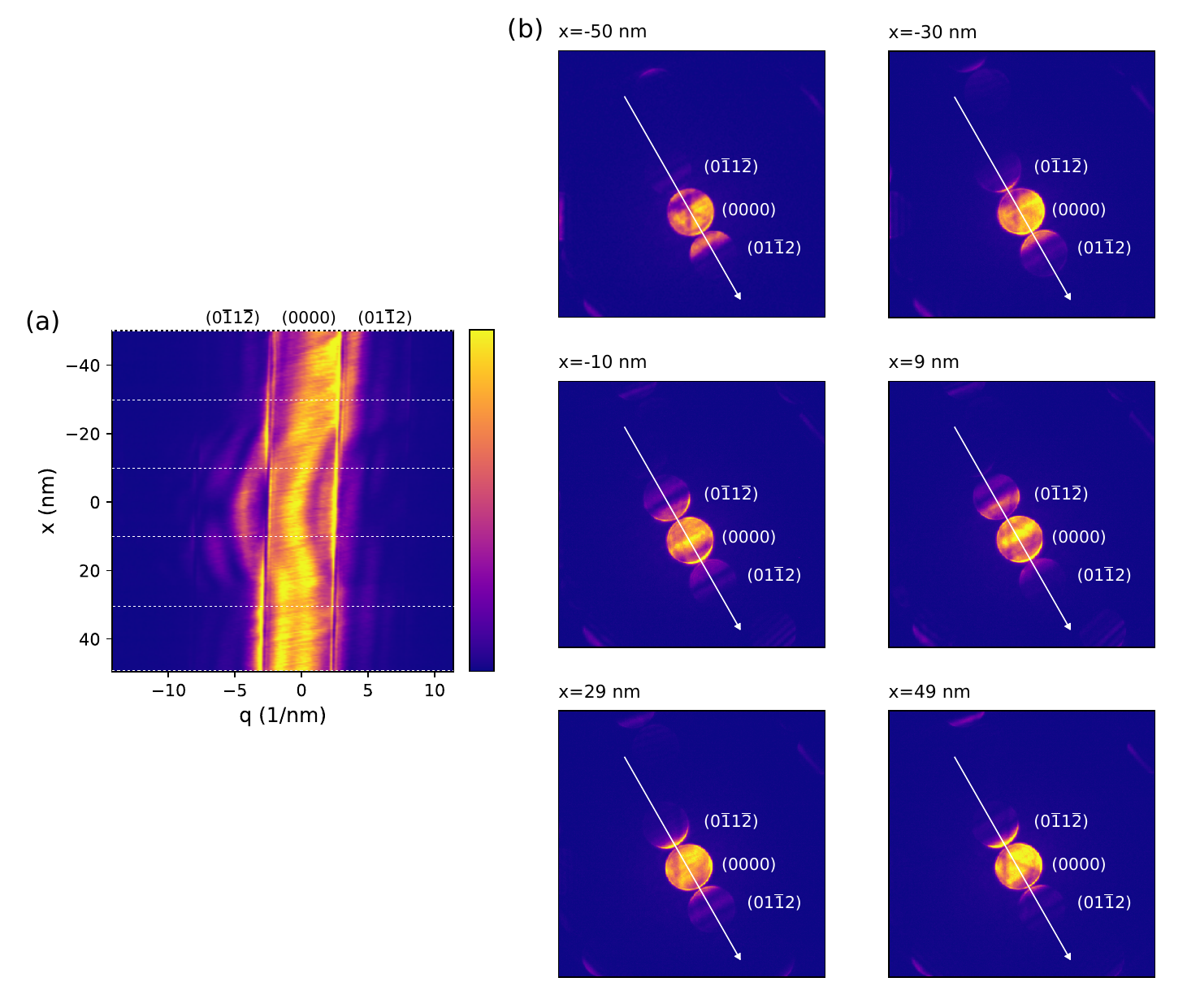}

\caption{Exemplary CBED patterns for several $x$-positions. (a) The $qx$-plots
with marked x-positions, for which the CBED patterns are shown. (b)
CBED patterns for the indicated $x$-positions. The arrow indicates,
where the profile was obtained. The intensities within the $qx$-plot
and the individual diffraction patterns are displayed on a common
color scale, which is given in (a)\label{fig:Examplary-CBED-patterns}}
\end{figure}

These diffraction patterns are shown for exemplary $x$-positions
in Fig.~\ref{fig:Examplary-CBED-patterns}. The dataset was further
reduced by obtaining profiles within the diffraction patterns along
the systematic row in the central part of the CBED disks. The arrows
in Fig.~\ref{fig:Examplary-CBED-patterns}(b) mark the position of
this profile. These profiles were averaged over 9~px in the direction
perpendicular to the arrow. This averaging was required to average
out the influence of the HOLZ-lines noticeable within the $(0000)$-diffraction
disk. 

After obtaining this second profile, a 2D dataset is left, which is
displayed within the $qx$-plot as shown in Fig.~\ref{fig:Examplary-CBED-patterns}(b)
(this is the same $qx$-plot as shown in Fig. 4(b) of the main text).
For each position $x$ within this plot the profile of the corresponding
diffraction pattern is shown along the $q$-position.

\section{Experimental and reconstruction parameters}

Table~\ref{tab:Parameters} gives the detailed parameters used in
the simulations, experiments and reconstructions.
\begin{table}[h]
\begin{tabular}{lccc}
 & Simulated & Simulated & Experimental\tabularnewline
Parameter & precipitate & inclined layers & inclined layer\tabularnewline
\hline 
\hline 
Semiconvergence angle (mrad) & 4.3 & 6.3 & 5.0\tabularnewline
Residual beam tilt (mrad) & 0.0 & 0.0 & 0.5\tabularnewline
$x$-sampling $qx$-plot (nm) & 2.0 & 2.0 & 0.31 (see text)\tabularnewline
$q$-sampling $qx$-plot (1/nm) & 0.1 & 0.1 & 0.057\tabularnewline
Regularization strength $\lambda$ & 125 & $2\times10^{4}$ & $3\times10^{9}$ and \tabularnewline
Max. half-angle $\Theta_{\text{max}}$ (mrad) & 4.3 & 6.3 & 4.4 (see text)\tabularnewline
Smoothing parameter $\beta$ (nm) & 20 & 20 & 10\tabularnewline
Grid spacing $\delta x$ (nm) & 2.56 & 6.74 & 7.14\tabularnewline
Grid spacing $\delta z$ (nm) & 2.1 & 10.45 & 6.87\tabularnewline
\hline 
\end{tabular}

\caption{Parameters in experiments and reconstruct\label{tab:Parameters}ions}
\end{table}

The $qx-$plots within the experimental inclined layer shows some
diffraction effects at the condenser aperture edges. For avoidance
of reconstructions artifacts that originate from these effects, a
parameter of $\Theta_{\mathrm{max}}=4.4\thinspace\text{mrad}$ was
used, which is slightly smaller than the experimental semiconvergence
angle. During the reconstruction of the strain field a coarser $x$-spacing
of only $2$~nm was used within the calculated $qx-$plots for efficiency.
For the evaluation of the residual norm these coarser points within
$I(x,q)$ were linearly interpolated in the $x$-direction. The $q$-spacing
was identical to the spacing in the experiment. 

\end{document}